# An Equivalence between Network Coding and Index Coding

M. Effros
California Institute of Technology
effros@caltech.edu

S. El Rouayheb
Princeton University
salim@princeton.edu

M. Langberg
The Open University of Israel
mikel@openu.ac.il

*Abstract*—We show that the network coding and index coding problems are equivalent. This equivalence holds in the general setting which includes *linear and non-linear* codes. Specifically, we present an efficient reduction that maps a network coding instance to an index coding one while preserving feasibility. Previous connections were restricted to the linear case.

## I. Introduction

In the network coding paradigm, a set of source nodes transmits information to a set of terminal nodes over a network while internal nodes of the network may mix received information before forwarding it. This mixing (or encoding) of information has been extensively studied over the last decade (see e.g., [1], [2], [3], [4], [5] and references therein). While network coding in the *multicast* setting is currently well understood, this is far from being the case for the general multi-source multi-terminal setting. In particular, determining the capacity of a general network coding instance remains an intriguing central open problem, e.g., [6], [7], [8], [9], [10].

A special instance to the network coding problem introduced in [11], which has seen significant interest lately, is the so-called *index coding* problem [11], [12], [13], [14], [15], [16]. Roughly speaking, the index coding problem encapsulates the "broadcast with side information" problem in which a single server wishes to communicate with several clients each requiring potentially different information and having potentially different side information (see Figure 1(a) for an example).

One may consider the index coding problem as a *simple* and *representative* instance of network coding. The instance is "simple" in the sense that any index coding instance can be represented as a topologically simple network coding instance in which only a *single* internal node has in-degree greater than one and thus only a single internal node can perform encoding (see Figure 1(b) for an example). It is "representative" in the sense that the index coding paradigm is broad enough to characterize the network coding problem under the assumption of *linear* encoding [17]. Specifically, given any instance to the network coding problem $\mathcal{I}$, one can efficiently construct an instance of the index coding problem $\hat{\mathcal{I}}$ such that: (a)

The work of Michael Langberg was supported in part by ISF grant 480/08 and BSF grant 2010075. The work of S. El Rouayheb was supported in part by the National Science Foundation under Grant CCF-1016671. Work done while Michael Langberg was visiting the California Institute of Technology. Authors appear in alphabetical order.

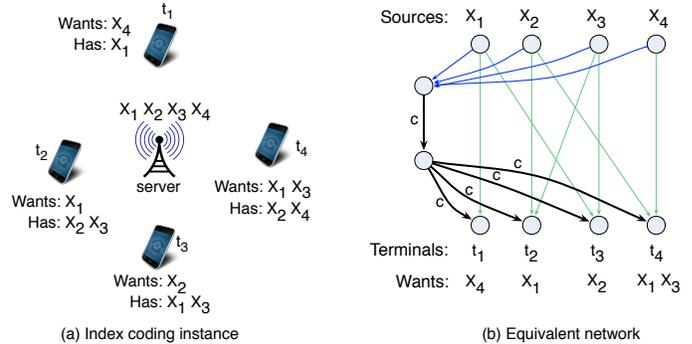

Fig. 1. (a) An instance of the index coding problem. A server has 4 binary sources $X_1, \ldots, X_4$ and there are 4 terminals with different "wants" and "has" sets (corresponding to the communication demand and side information respectively). How can we satisfy the terminals' demands with a minimum number of bits broadcasted by the server? The server can trivially transmit all the 4 sources. However, this is not optimal and it is sufficient to broadcast only 2 bits, namely $X_1 + X_2 + X_3$ and $X_1 + X_4$ ('+' denotes the xor operation). (b) Index coding is a special case of the network coding problem. All links are of unit capacity (non-specified) or of capacity $c$. Links directly connecting between sources and terminals represent the "has" sets. Any solution to the index coding problem with $c$ broadcast bits can be efficiently mapped to a solution to the corresponding network coding instance and visa versa. This implies that the index coding problem is a special case of the network coding problem. The focus of this work is on the opposite assertion. Namely, that the network coding problem is a special case of the index coding problem.

There exists a linear solution to $\mathcal{I}$ if and only if there exists an optimal linear solution to $\hat{\mathcal{I}}$, and (b) any optimal linear solution to $\hat{\mathcal{I}}$ can be efficiently turned into a linear solution to $\mathcal{I}$. All undefined notions above (and those that follow), such as "solution," "feasibility," and "capacity", will be given in detail in Section II.

The results of [17] hold for (scalar and vector) linear coding functions only, and the analysis there breaks down once one allows general coding (which may be non-linear) at internal nodes. The study of non-linear coding functions is central to the study of network coding as it is shown in [18] that non-linear codes have an advantage over linear solutions, *i.e.*, that there exist instances in which linear codes do not suffice to achieve capacity.

In this work, we extend the equivalence between network coding and index coding to the setting of general encoding functions (which may be non-linear). Our results imply that, effectively, when one wishes to solve a network coding instance $\mathcal{I}$, a possible route is to turn the network coding instance into an index coding instance $\hat{\mathcal{I}}$ (via our reduction), solve the index coding instance $\hat{\mathcal{I}}$, and turn the solution to $\hat{\mathcal{I}}$ into a solution to the original network coding instance $\mathcal{I}$. Hence, any efficient scheme to solve index coding will yield an efficient scheme for network coding. Stated differently, our results imply that an understanding of the solvability of index coding instances will imply an understanding of the solvability of network coding instances as well.

The remainder of the paper is structured as follows. In Section II, we present the models of network and index coding. In Section III, we present an example based on the "butterfly network" that illustrates our proof techniques. In Section IV, we present the main technical contribution of this work: the equivalence between network and index coding. In Section V, we show a connection between the capacity regions of index coding and network coding in networks with collocated sources. Finally, in Section VI, we conclude with some remarks and open problems.

## II. MODEL

In what follows we define the model for the network coding and index coding problems. Throughout this paper, "hatted" variables (e.g., $\hat{x}$) will correspond to the variables of index coding instances, while "unhatted" variables will correspond to the network coding instance. For integers $k > 0$, we use $[k]$ to denote the set $\{1, \ldots, k\}$.

### A. Network coding

An instance $\mathcal{I} = (G, S, T, B)$ of the network coding problem includes a directed acyclic network $G = (V, E)$, a set of sources nodes $S \subset V$, a set of terminal nodes $T \subset V$, and an $|S| \times |T|$ requirement matrix $B$. We assume, without loss of generality, that each source $s \in S$ has no incoming edges and that each terminal $t \in T$ has no outgoing edges. Let $c_e$ denote the capacity of each edge $e \in E$, namely for any block length $n$, each edge $e$ can carry one of the $2^{c_e n}$ messages in $[2^{c_e n}]$. In our setting, each source $s \in S$ holds a rate $R_s$ random variable $X_s$ uniformly distributed over $[2^{R_s n}]$. The variables describing different messages are independent. We assume that values of the form $2^{c_e n}$ and $2^{R_s n}$ are integers.

A network code, $(\mathcal{F}, \mathcal{X}) = \{(f_e, X_e)\} \cup \{g_t\}$, is an assignment of a pair $(f_e, X_e)$ to each edge $e \in E$ and a decoding function $\{g_t\}$ to each terminal $t \in T$. For $e = (u, v)$, $f_e$ is a function taking as input the random variables associated with incoming edges to node $u$, and $X_e \in [2^{c_e n}]$ is the random variable equal to the evaluation of $f_e$ on its inputs. If $e$ is an edge leaving a source node $s \in S$, then $X_s$ is the input to $f_e$. The input to the decoding function $g_t$ consists of the random variables associated with incoming edges to terminal $t$. The output of $g_t$ is required to be a vector of all sources required by $t$.

Given the acyclic structure of $G$, the functions $\{f_e\}$ and their evaluation $\{X_e\}$ can be defined by induction on the topological order of $G$. Namely, given the family $\{f_e\}$ one can define a function family $\{\bar{f}_e\}$[1] such that each $\bar{f}_e$ takes as input the source information $\{X_s\}$ and outputs the random variable $X_e$. More precisely, for $e = (u, v)$ in which $u$ is a source node, define $\bar{f}_e = f_e$. For $e = (u, v)$ in which $u$ is an internal node with incoming edges $In(e) = \{e'_1, \ldots, e'_\ell\}$, define $\bar{f}_e = f_e(\bar{f}_{e'_1}, \ldots, \bar{f}_{e'_\ell})$. Namely, the evaluation of $\bar{f}_e$ on source information $\{X_s\}$ equals the evaluation of $f_e$ given the values of $\bar{f}_{e'}$ for $e' \in In(e)$. We will use both $\{f_e\}$ and $\{\bar{f}_e\}$ in our analysis.

The $|S| \times |T|$ requirement matrix $B = [b_{i,j}]$ has entries in the set $\{0, 1\}$, with $b_{s,t} = 1$ if and only if terminal $t$ requires information from source $s$.

A network code $(\mathcal{F}, \mathcal{X})$ is said to satisfy terminal node $t$ under transmission $(x_s : s \in S)$ if the decoding function $g_t$ outputs $(x_s : b(s, t) = 1)$ when $(X_s : s \in S) = (x_s : s \in S)$. The Network code $(\mathcal{F}, \mathcal{X})$ is said to satisfy the instance $\mathcal{I}$ with error probability $\varepsilon \geq 0$ if the probability that all $t \in T$ are simultaneously satisfied is at least $1 - \varepsilon$. The probability is taken over the joint distribution on random variables $(X_s : s \in S)$.

For a rate tuple $R = (R_1, \ldots, R_{|S|})$, an instance $\mathcal{I}$ to the network coding problem is said to be $(\varepsilon, R, n)$-feasible if there exists a network code $(\mathcal{F}, \mathcal{X})$ with block length $n$ that satisfies $\mathcal{I}$ with error at most $\varepsilon$ when applied to source information $(X_1, \ldots, X_{|S|})$, where $X_s$ is uniformly distributed over $[2^{R_s n}]$. An instance $\mathcal{I}$ to the network coding problem is said to be $R$-feasible if for any $\varepsilon > 0$ and any $\delta > 0$ there exists a block length $n$ such that $\mathcal{I}$ is $(\varepsilon, R(1-\delta), n)$-feasible. Here, $R(1-\delta) = (R_1(1-\delta), \ldots, R_{|S|}(1-\delta))$. The capacity region of an instance $\mathcal{I}$ refers to all rate tuples $R$ for which $\mathcal{I}$ is $R$-feasible.

### B. Index coding

The index coding problem captures the "broadcast with side information" problem in which a single server wishes to communicate with several clients each having different side information. Specifically, an instance to index coding includes a set of terminals $\hat{T} = \{\hat{t}_1, \ldots, \hat{t}_{|\hat{T}|}\}$ and a set of sources $\hat{S} = \{\hat{s}_1, \hat{s}_2, \ldots, \hat{s}_{|\hat{S}|}\}$ available at the server. Given a block length $n$, source $\hat{s}$ holds a rate $\hat{R}_{\hat{s}}$ random variable $\hat{X}_{\hat{s}}$ uniformly distributed over $[2^{\hat{R}_{\hat{s}} n}]$ (and independent from other sources). Each terminal requires information from a certain subset of sources in $\hat{S}$. In addition, information from some sources in $\hat{S}$ are available a priori as side information to each terminal. Specifically, terminal $\hat{t} \in \hat{T}$ is associated with sets:

- $\hat{W}_{\hat{t}}$ is the set of sources required by $\hat{t}$, and
- $\hat{H}_{\hat{t}}$ is the set of sources available at $\hat{t}$.

We refer to $\hat{W}_{\hat{t}}$ and $\hat{H}_{\hat{t}}$ as the "wants" and "has" sets of $\hat{t}$, respectively. The server uses an error-free broadcast channel to transmit information to the terminals. The objective is to design an encoding scheme that satisfy the demands of all the terminals while minimizing the number of uses of the broadcast channel (See Figure 1).

Formally, an instance $\hat{\mathcal{I}}$ to the index coding problem is the tuple $(\hat{S}, \hat{T}, \{\hat{W}_{\hat{t}}\}, \{\hat{H}_{\hat{t}}\})$. An index code $(\hat{\mathcal{E}}_B, \hat{\mathcal{D}})$ for $\hat{\mathcal{I}}$ with broadcast rate $\hat{c}_B$, includes an encoding function $\hat{\mathcal{E}}_B$ for the broadcast channel, and a set of decoding functions $\hat{\mathcal{D}} = \{\hat{D}_{\hat{t}}\}_{\hat{t} \in \hat{T}}$ with one function for each terminal. The function $\hat{\mathcal{E}}_B$ is a function that takes as input the source random variables $\{\hat{X}_{\hat{s}}\}$ and outputs a rate $\hat{c}_B$ random variable $\hat{X}_B \in [2^{\hat{c}_B n}]$. The input to the decoding function $\hat{D}_{\hat{t}}$ consists of the random variables in $\hat{H}_{\hat{t}}$ (the source random variables available to $\hat{t}$) and the broadcast message $\hat{X}_B$. The output of $\hat{D}_{\hat{t}}$ is intended to be a vector of all sources in $\hat{W}_{\hat{t}}$ required by $\hat{t}$.

---
[1] In the network coding literature, $\{f_e\}$ and $\{\bar{f}_e\}$ are sometimes referred to as the local and global encoding functions, respectively.

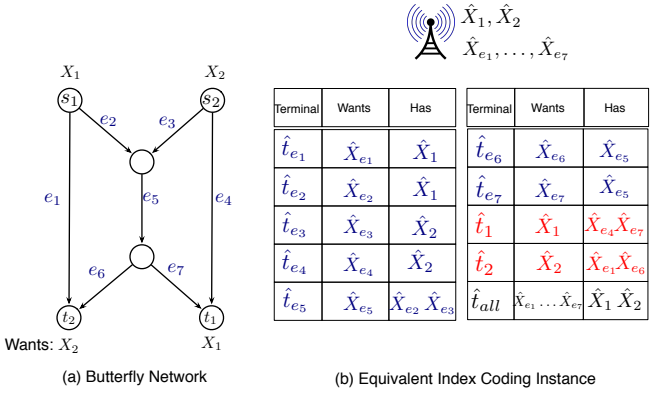

Fig. 2. (a) The butterfly network with two sources $X_1$ and $X_2$ and two terminals $t_1$ and $t_2$. (b) The equivalent index coding instance. The server has 9 sources: one for each source, namely $\{\hat{X}_1, \hat{X}_2\}$, and one for each edge in the network, namely $\{\hat{X}_{e_1}, \ldots, \hat{X}_{e_7}\}$. There are 7 clients corresponding to the 7 edges in the network, 2 clients corresponding to the two terminals of the butterfly network and one extra terminal $\hat{t}_{all}$.

An index code $(\hat{\mathcal{E}}_B, \hat{\mathcal{D}})$ of broadcast rate $\hat{c}_B$ is said to satisfy terminal $\hat{t}$ under transmission $(\hat{x}_{\hat{s}} : \hat{s} \in \hat{S})$ if the decoding function $\hat{D}_{\hat{t}}$ outputs $(\hat{x}_{\hat{s}} : \hat{s} \in \hat{W}_{\hat{t}})$ when $(\hat{X}_{\hat{s}} : \hat{s} \in \hat{S}) = (\hat{x}_{\hat{s}} : \hat{s} \in \hat{S})$. Index code $(\hat{\mathcal{E}}_B, \hat{\mathcal{D}})$ is said to satisfy instance $\hat{\mathcal{I}}$ with error probability $\varepsilon \geq 0$ if the probability that all $\hat{t} \in \hat{T}$ are simultaneously satisfied is at least $1 - \varepsilon$. The probability is taken over the joint distribution on random variables $\{\hat{X}_{\hat{s}}\}_{\hat{s} \in \hat{S}}$.

For a rate tuple $\hat{R} = (\hat{R}_1, \ldots, \hat{R}_{|\hat{S}|})$ and broadcast rate $\hat{c}_B$, an instance $\hat{\mathcal{I}}$ to the index coding problem is said to be $(\varepsilon, \hat{R}, \hat{c}_B, n)$-feasible if there exists an index code $(\hat{\mathcal{E}}_B, \hat{\mathcal{D}})$ with broadcast rate $\hat{c}_B$ and block length $n$ that satisfies $\hat{\mathcal{I}}$ with error at most $\varepsilon$ when applied to source information $(\hat{X}_{\hat{s}} : \hat{s} \in \hat{S})$ where each $\hat{X}_{\hat{s}}$ is uniformly and independently distributed over $[2^{\hat{R}_{\hat{s}} n}]$. An instance $\hat{\mathcal{I}}$ to the network coding problem is said to be $(\hat{R}, \hat{c}_B)$-feasible if for any $\varepsilon > 0$ and $\delta > 0$ there exists a block length $n$ such that $\hat{\mathcal{I}}$ is $(\varepsilon, \hat{R}(1-\delta), \hat{c}_B, n)$-feasible. As before, $\hat{R}(1-\delta) = (\hat{R}_1(1-\delta), \ldots, \hat{R}_{|\hat{S}|}(1-\delta))$. The capacity region of an instance $\hat{\mathcal{I}}$ refers to all rate tuples $\hat{R}$ and capacities $\hat{c}_B$ for which $\hat{\mathcal{I}}$ is $(\hat{R}, \hat{c}_B)$-feasible.

## III. EXAMPLE

Our main result states that the network coding and index coding problems are equivalent (for linear and non-linear coding). Theorem 1 in Section IV gives a formal statement of this result. The proof is based on a reduction that constructs for any given network coding problem an equivalent index coding problem. In this section, we explain the main elements of our proof by applying it to the butterfly network [1] example in Fig. 2(a). For simplicity, our example does not consider any error in communication. Our reduction goes along the lines of the construction in [17], while our analysis differs to capture the case of non-linear encoding.

We start by briefly describing the butterfly network depicted in Fig. 2(a). The network has two information sources $s_1$ and $s_2$ that hold two uniform i.i.d binary random variables $X_1$ and $X_2$, respectively. There are also two terminals (destinations) $t_1$ and $t_2$ that want $X_1$ and $X_2$, respectively. All the edges in the network, $e_1, \ldots, e_7$, have capacity 1. Following the notation in Section II-A, let $\bar{f}_{e_i}(X_1, X_2)$ be the one-bit message on edge $e_i$. Then, the following is a network code that satisfies the demands of the terminals:

$$\begin{aligned}\bar{f}_{e_1} &= \bar{f}_{e_2} = X_1 \\ \bar{f}_{e_3} &= \bar{f}_{e_4} = X_2 \\ \bar{f}_{e_5} &= \bar{f}_{e_6} = \bar{f}_{e_7} = X_1 + X_2,\end{aligned} \quad (1)$$

where '+' denotes the xor operation. Terminal $t_1$ can decode $X_1$ by computing $X_1 = \bar{f}_{e_4} + \bar{f}_{e_7}$, and $t_2$ can decode $X_2$ by computing $X_2 = \bar{f}_{e_1} + \bar{f}_{e_6}$. Thus, the butterfly network is $(\epsilon, R, 1) = (0, (1, 1), 1)$-feasible.

The problem now is to construct an index coding instance that is "equivalent" to the butterfly network, i.e., any index code for that instance would imply a network code for the butterfly network, and vice versa. We propose the following construction, based on that presented in [17], in which the server has 9 sources split into two sets (see Figure 2).

- $\hat{X}_1$ and $\hat{X}_2$ corresponding to the two sources $X_1$ and $X_2$ in the butterfly network.
- $\hat{X}_{e_1}, \ldots, \hat{X}_{e_7}$ corresponding to the edges $e_1, \ldots, e_7$ in the butterfly network.

There are 10 clients, as described in the Table in Fig. 2(b). These clients are split into 3 sets:

- A client $\hat{t}_{e_i}$ for each edge $e_i$. Client $\hat{t}_{e_i}$ wants $\hat{X}_{e_i}$ and has the variables $\hat{X}_{e_j}$ for each edge $e_j$ in the butterfly network that is incoming to $e_i$.
- A client $\hat{t}_i$ for each network terminal $t_i$. $\hat{t}_i$ wants $\hat{X}_i$ and has the variables $\hat{X}_{e_j}$ for each edge $e_j$ in the butterfly network that is incoming to $t_i$. Namely $\hat{t}_1$ wants $\hat{X}_1$ and has $\hat{X}_{e_4}$ and $\hat{X}_{e_7}$, whereas $\hat{t}_2$ wants $\hat{X}_2$ and has $\hat{X}_{e_1}$ and $\hat{X}_{e_6}$.
- One client $\hat{t}_{all}$ that wants all variables that correspond to edges of the butterfly network (i.e., $\hat{X}_{e_1}, \ldots, \hat{X}_{e_7}$) and has all variables that correspond to sources of the butterfly network (i.e., $\hat{X}_1$ and $\hat{X}_2$).

Next, we explain how the solutions are mapped between these two instances. While "Direction 1" strongly follows the analysis appearing in [17], our major novelty is in "Direction 2" (both proof directions are presented below for completion).

*Direction 1: Network code to index code.* Suppose we are given a network code with local encoding functions $f_{e_i}$, and global encoding functions $\bar{f}_{e_i}(X_1, X_2), i = 1, \ldots, 7$. We construct the following index code solution in which the server broadcasts the 7-bit vector $\hat{X}_B = (\hat{X}_B(e_1), \ldots, \hat{X}_B(e_7))$,

where
$$\hat{X}_B(e_i) = \hat{X}_{e_i} + \bar{f}_{e_i}(\hat{X}_1, \hat{X}_2), \quad i = 1, \ldots, 7. \quad (2)$$

For instance, the index code corresponding to the network code in (1) is
$$\begin{aligned}
\hat{X}_B(e_1) &= \hat{X}_{e_1} + \hat{X}_1 \\
\hat{X}_B(e_2) &= \hat{X}_{e_2} + \hat{X}_1 \\
\hat{X}_B(e_3) &= \hat{X}_{e_3} + \hat{X}_2 \\
\hat{X}_B(e_4) &= \hat{X}_{e_4} + \hat{X}_2 \\
\hat{X}_B(e_5) &= \hat{X}_{e_5} + \hat{X}_1 + \hat{X}_2 \\
\hat{X}_B(e_6) &= \hat{X}_{e_6} + \hat{X}_1 + \hat{X}_2 \\
\hat{X}_B(e_7) &= \hat{X}_{e_7} + \hat{X}_1 + \hat{X}_2.
\end{aligned} \quad (3)$$

One can check that this index code allows each client to recover the sources in its "wants" set using the broadcast $\hat{X}_B$ and the information in its "has" set. For example, in the case of the index code in (3), terminal $\hat{t}_{e_5}$ computes
$$\hat{X}_{e_5} = \hat{X}_B(e_5) - (\hat{X}_B(e_2) - \hat{X}_{e_2}) - (\hat{X}_B(e_3) - \hat{X}_{e_3}).$$

Here, both '+' and '-' denote the xor operation. More specifically, terminal $\hat{t}_{e_5}$ first computes $\bar{f}_{e'}$ for its incoming edges via its "has" set and $\hat{X}_B$ (i.e., $\bar{f}_{e_2} = \hat{X}_B(e_2) - \hat{X}_{e_2}$ and $\bar{f}_{e_3} = \hat{X}_B(e_3) - \hat{X}_{e_3}$). Then using the fact that
$$\bar{f}_{e_5}(\hat{X}_1, \hat{X}_2) = f_{e_5}\left(\bar{f}_{e'}(\hat{X}_1, \hat{X}_2) | e' \text{ is an incoming edge of } e_5\right),$$

terminal $\hat{t}_{e_5}$ can compute $\bar{f}_{e_5}(\hat{X}_1, \hat{X}_2)$. Finally, by the definition of $\hat{X}_B$ in (2) terminal $\hat{t}_{e_5}$ recovers $\hat{X}_{e_5} = \hat{X}_B(e_5) - \bar{f}_{e_5}(\hat{X}_1, \hat{X}_2)$. By a similar process, every terminal in the index coding instance can decode the sources it wants.

*Direction 2: Index code to network code.* Let $\hat{c}_B$ equal the total capacity of edges in the butterfly network, i.e., $\hat{c}_B = 7$. Suppose we are given an index code with broadcast rate $\hat{c}_B$ that allows each client to decode the sources it requires (with no errors). We want to show that any such code can be mapped to a network code for the butterfly network. Let us denote by $\hat{X}_B = (\hat{X}_{B,1}, \ldots, \hat{X}_{B,7})$ the broadcast information where $\hat{X}_B$ is a function, possibly non-linear, of the 9 sources available at the server $\hat{X}_1, \hat{X}_2$ and $\hat{X}_{e_1}, \ldots, \hat{X}_{e_7}$.

For every client $\hat{t}$, there exists a decoding function $\hat{D}_{\hat{t}}$ that takes as input the broadcast information $\hat{X}_B$ and the sources in its "has" set and outputs the sources it requires. Namely, we have the following functions:
$$\begin{aligned}
\hat{D}_{\hat{t}_{e_1}}(\hat{X}_B, \hat{X}_1) &= \hat{X}_{e_1} & \hat{D}_{\hat{t}_{e_2}}(\hat{X}_B, \hat{X}_1) &= \hat{X}_{e_2} \\
\hat{D}_{\hat{t}_{e_3}}(\hat{X}_B, \hat{X}_2) &= \hat{X}_{e_3} & \hat{D}_{\hat{t}_{e_4}}(\hat{X}_B, \hat{X}_2) &= \hat{X}_{e_4} \\
\hat{D}_{\hat{t}_{e_5}}(\hat{X}_B, \hat{X}_{e_2}, \hat{X}_{e_3}) &= \hat{X}_{e_5} & \hat{D}_{\hat{t}_{e_6}}(\hat{X}_B, \hat{X}_{e_5}) &= \hat{X}_{e_6} \quad (4) \\
\hat{D}_{\hat{t}_{e_7}}(\hat{X}_B, \hat{X}_{e_5}) &= \hat{X}_{e_7} & \hat{D}_{\hat{t}_1}(\hat{X}_B, \hat{X}_{e_4}, \hat{X}_{e_7}) &= \hat{X}_1 \\
\hat{D}_{\hat{t}_2}(\hat{X}_B, \hat{X}_{e_1}, \hat{X}_{e_6}) &= \hat{X}_2 & \hat{D}_{\hat{t}_{all}}(\hat{X}_B, \hat{X}_1, \hat{X}_2) & \\
& & &= (\hat{X}_{e_1}, \ldots, \hat{X}_{e_7}).
\end{aligned}$$

We will use these decoding functions to construct the network code for the butterfly network. Consider for example edge $e_5$. Its incoming edges are $e_2$ and $e_3$, so we need to define a local encoding $f_{e_5}$ which is a function of the information $X_{e_2}$ and $X_{e_3}$ they are carrying. In our approach, we fix a specific value $\sigma$ for $\hat{X}_B$, and define
$$f_{e_5}(X_{e_2}, X_{e_3}) = \hat{D}_{\hat{t}_{e_5}}(\sigma, X_{e_2}, X_{e_3}).$$

Similarly, we define the encoding functions for every edge in the butterfly network, and the decoding functions for the two terminals $t_1$ and $t_2$. The crux of our proof lies in showing that there exists a value of $\sigma$ for which the corresponding network code allows correct decoding. In the example at hand, one may choose $\sigma$ to be the all zero vector $\mathbf{0}$ (or actually any vector for that matter). The resulting network code is:
$$\begin{aligned}
f_{e_1} &= \hat{D}_{\hat{t}_{e_1}}(\mathbf{0}, X_1) & f_{e_2} &= \hat{D}_{\hat{t}_{e_2}}(\mathbf{0}, X_1) \\
f_{e_3} &= \hat{D}_{\hat{t}_{e_3}}(\mathbf{0}, X_2) & f_{e_4} &= \hat{D}_{\hat{t}_{e_4}}(\mathbf{0}, X_2) \quad (5) \\
f_{e_5} &= \hat{D}_{\hat{t}_{e_5}}(\mathbf{0}, f_{e_2}, f_{e_3}) & f_{e_6} &= \hat{D}_{\hat{t}_{e_6}}(\mathbf{0}, f_{e_5}) \\
f_{e_7} &= \hat{D}_{\hat{t}_{e_7}}(\mathbf{0}, f_{e_5}). &
\end{aligned}$$

Terminals $t_1$ and $t_2$ can decode using the functions $\hat{D}_{\hat{t}_1}(\mathbf{0}, f_{e_4}, f_{e_7})$ and $\hat{D}_{\hat{t}_2}(\mathbf{0}, f_{e_1}, f_{e_6})$, respectively.

To prove correct decoding, we show that for any fixed values of $\hat{X}_1$ and $\hat{X}_2$, there exists a unique value for the vector $(\hat{X}_{e_1}, \ldots, \hat{X}_{e_7})$ that corresponds to $\hat{X}_B = \mathbf{0}$. Otherwise, it can be seen that $\hat{t}_{all}$ cannot decode correctly since $\hat{c}_B = 7$ and $\hat{X}_B$ is a function of $\hat{X}_1, \hat{X}_2$ and $\hat{X}_{e_1}, \ldots, \hat{X}_{e_7}$. Roughly speaking, this correspondence allows us to reduce the analysis of correct decoding in the resulting network code, to correct decoding in the original index code. Full details of this reduction, and on how to choose $\sigma$ appear in the upcoming Section IV.

## IV. MAIN RESULT

We follow the proof of [17] to obtain our main result.

*Theorem 1:* For any instance to the network coding problem $\mathcal{I}$ one can efficiently construct an instance to the index coding problem $\hat{\mathcal{I}}$ and an integer $\hat{c}_B$ such that for any rate tuple $R$, any integer $n$, and any $\varepsilon \geq 0$ it holds that $\mathcal{I}$ is $(\varepsilon, R, n)$ feasible iff $\hat{\mathcal{I}}$ is $(\varepsilon, \hat{R}, \hat{c}_B, n)$ feasible. Here, the rate vector $\hat{R}$ for $\hat{\mathcal{I}}$ can be efficiently computed from $R$; and the corresponding network and index codes that imply feasibility in the reduction can be efficiently constructed from one another.

*Proof:* Let $G = (V, E)$, and $\mathcal{I} = (G, S, T, B)$. Let $n$ be any integer, and let $R = (R_1, \ldots, R_{|S|})$. We start by defining $\hat{\mathcal{I}} = (\hat{S}, \hat{T}, \{\hat{W}_{\hat{t}}\}, \{\hat{H}_{\hat{t}}\})$, the integer $\hat{c}_B$, and the rate tuple $\hat{R}$. See Figure 2 for an example. To simplify notation, we use the notation $\hat{X}_{\hat{s}}$ to denote both the source $\hat{s} \in \hat{S}$ and the corresponding random variable. For $e = (u, v)$ in $E$ let $In(e)$ be the set of edges entering $u$ in $G$. If $u$ is a source $s$ let $In(e) = \{s\}$. For $t_i \in T$, let $In(t_i)$ be the set of edges entering $t_i$ in $G$.

- $\hat{S}$ consists of $|S| + |E|$ sources: one source denoted $\hat{X}_s$ for each original source $s$ in $\mathcal{I}$ and one source denoted $\hat{X}_e$ for each edge $e$ in $G$. Namely, $\hat{S} = \{\hat{X}_s\}_{s \in S} \cup \{\hat{X}_e\}_{e \in E}$.
- $\hat{T}$ consists of $|E| + |T| + 1$ terminals: $|E|$ terminals denoted $\hat{t}_e$ corresponding to the edges in $E$, $|T|$ terminals

denoted $\hat{t}_i$ corresponding to the terminals in $\mathcal{I}$, and a single terminal denoted $\hat{t}_{all}$. Namely, $\hat{T} = \{\hat{t}_e\}_{e \in E} \cup \{\hat{t}_i\}_{i \in [|T|]} \cup \{\hat{t}_{all}\}$.

- For $\hat{t}_e \in \hat{T}$ we set $\hat{H}_{\hat{t}_e} = \{\hat{X}_{e'}\}_{e' \in In(e)}$ and $\hat{W}_{\hat{t}_e} = \{\hat{X}_e\}$.
- For $\hat{t}_i \in \hat{T}$, let $t_i$ be the corresponding terminal in $T$. We set $\hat{H}_{\hat{t}_i} = \{\hat{X}_{e'}\}_{e' \in In(t_i)}$ and $\hat{W}_{\hat{t}_i} = \{\hat{X}_s\}_{s:b(s,t_i)=1}$.
- For $\hat{t}_{all}$ set $\hat{H}_{\hat{t}_{all}} = \{\hat{X}_s\}_{s \in S}$ and $\hat{W}_{\hat{t}_{all}} = \{\hat{X}_e\}_{e \in E}$.
- Let $\hat{R}$ be a vector of length $|S| + |E|$ consisting of two parts: $(\hat{R}_s : s \in S)$ represents the rate $\hat{R}_s$ of each $\hat{X}_s$ and $(\hat{R}_e : e \in E)$ represents the rate $\hat{R}_e$ of $\hat{X}_e$. Set $\hat{R}_s = R_s$ for each $s \in S$ and $\hat{R}_e = c_e$ for each $e \in E$. (Here $R_s$ is the entry corresponding to $s$ in the tuple $R$, and $c_e$ is the capacity of the edge $e$ in $G$.)
- Set $\hat{c}_B$ to be equal to $\sum_{e \in E} c_e = \sum_{e \in E} \hat{R}_e$.

We now present the two directions of our proof. The fact that $\mathcal{I}$ is $(\varepsilon, R, n)$ feasible implies that $\hat{\mathcal{I}}$ is $(\varepsilon, \hat{R}, \hat{c}_B, n)$ feasible was essentially shown in [17] and is presented here for completeness. The other direction is the major technical contribution of this work.

**$\mathcal{I}$ is $(\varepsilon, R, n)$ feasible implies that $\hat{\mathcal{I}}$ is $(\varepsilon, \hat{R}, \hat{c}_B, n)$:**

For this direction we assume the existence of a network code $(\mathcal{F}, \mathcal{X}) = \{(f_e, X_e)\} \cup \{g_t\}$ for $\mathcal{I}$ which is $(\varepsilon, R, n)$ feasible. As mentioned in Section I, given the acyclic structure of $G$, one may define a new set of functions $\bar{f}_e$ with input $\{X_s\}_{s \in S}$ such that the evaluation of $\bar{f}_e$ is identical to the evaluation of $f_e$, which is $X_e$. We construct an index code $(\hat{\mathcal{E}}_B, \hat{\mathcal{D}}) = (\hat{\mathcal{E}}_B, \{\hat{D}_{\hat{t}}\})$ for $\hat{\mathcal{I}}$. We do this by specifying the broadcast encoding $\hat{\mathcal{E}}_B$ and the decoding functions $\{\hat{D}_{\hat{t}}\}_{\hat{t} \in \hat{T}}$.

The function $\hat{\mathcal{E}}_B$ will be defined in chunks, with one chunk (of support size $[2^{c_e n}]$) for each edge $e \in E$ denoted $\hat{\mathcal{E}}_B(e)$. We denote the output of $\hat{\mathcal{E}}_B(e)$ by $\hat{X}_B(e)$ and the output of $\hat{\mathcal{E}}_B$ by the concatenation $\hat{X}_B$ of the output chunks $\hat{X}_B(e)$. In what follows we use 'a+b' as the bitwise xor operator between equal length bit vectors $a$ and $b$. For each $e \in E$, the corresponding chunk in $\hat{\mathcal{E}}_B(e)$ will be equal to $\hat{X}_e + \bar{f}_e(\hat{X}_1, \ldots, \hat{X}_{|S|})$. It follows that $\hat{\mathcal{E}}_B$ is a function from the source random variables of $\hat{\mathcal{I}}$ to $\hat{X}_B$ with support

$$\left[2^{\sum_{e \in E} \hat{R}_e n}\right] = [2^{\hat{c}_B n}].$$

We now set the decoding functions:

- For $\hat{t}_e$ in $\hat{T}$ we set $\hat{D}_{\hat{t}_e}$ to be the function defined by the following decoding scheme:
  - First, for each $e' \in In(e)$, using the information in $\hat{H}_{\hat{t}_e}$, the decoder computes $\hat{X}_B(e') + \hat{X}_{e'} = \bar{f}_{e'}(\hat{X}_1, \ldots, \hat{X}_{|S|}) + \hat{X}_{e'} + \hat{X}_{e'} = \bar{f}_{e'}(\hat{X}_1, \ldots, \hat{X}_{|S|})$.
  - Then, let $In(e) = \{e'_1, \ldots, e'_\ell\}$. Using the function $f_e$ from network code $(\mathcal{F}, \mathcal{X})$, the decoder computes

    $$f_e(\bar{f}_{e'_1}(\hat{X}_1, \ldots, \hat{X}_{|S|}), \ldots, \bar{f}_{e'_\ell}(\hat{X}_1, \ldots, \hat{X}_{|S|})).$$

    By definition of $\bar{f}_e$ this is exactly $\bar{f}_e(\hat{X}_1, \ldots, \hat{X}_{|S|})$.
  - Finally, compute

    $$\hat{X}_B(e) + \bar{f}_e(\hat{X}_1, \ldots, \hat{X}_{|S|}) = \hat{X}_e,$$

    which is the source information client $\hat{t}_e$ wants in $\hat{\mathcal{I}}$.

- For $\hat{t}_i \in \hat{T}$ the process is almost identical to that above. Let $t_i$ be the corresponding terminal in $T$. The function $g_{t_i}$ is used on the evaluations of $\bar{f}_{e'}$ for $e' \in In(t_i)$, and the outcome is exactly the set of sources $\hat{W}_{\hat{t}_i} = \{\hat{X}_s\}_{s:b(s,t_i)=1}$ wanted by $t_i$.

- For $\hat{t}_{all}$, recall that $\hat{H}_{\hat{t}_{all}} = \{\hat{X}_s\}_{s \in S}$ and $\hat{W}_{\hat{t}_{all}} = \{\hat{X}_e\}_{e \in E}$. To obtain $\hat{X}_e$ the decoder evaluates $\hat{X}_B(e) + \bar{f}_e(\hat{X}_1, \ldots, \hat{X}_{|S|})$.

Let $\varepsilon \geq 0$. We now show that if the network code $(\mathcal{F}, \mathcal{X})$ succeeds with probability $1-\varepsilon$ on network $\mathcal{I}$ (over the sources $\{X_s\}_{s \in S}$ of rate tuple $R$), then the corresponding index code also succeeds with probability $1-\varepsilon$ over $\hat{\mathcal{I}}$ with sources $\{\hat{X}\}$ of rate tuple $\hat{R}$.

Consider any realization $\mathbf{x} = \{x_s\}$ of source information $\{X_s\}$ of the given network coding instance $\mathcal{I}$ for which all terminals of the network code decode successfully. Denote a realization of source information $\{\hat{X}_{\hat{s}}\}_{\hat{s} \in \hat{S}}$ in $\hat{\mathcal{I}}$ by $(\hat{\mathbf{x}}_\mathbf{s}, \hat{\mathbf{x}}_\mathbf{e})$, where $\hat{\mathbf{x}}_\mathbf{s}$ corresponds to the sources $\{\hat{X}_s\}_{s \in S}$ and $\hat{\mathbf{x}}_\mathbf{e}$ corresponds to sources $\{\hat{X}_e\}_{e \in E}$. Let $\hat{\mathbf{x}}_\mathbf{s}(s)$ be the entry in $\hat{\mathbf{x}}_\mathbf{s}$ corresponding to source $\hat{X}_s$ for $s \in S$, and let $\hat{\mathbf{x}}_\mathbf{e}(e)$ be the entry in $\hat{\mathbf{x}}_\mathbf{e}$ corresponding to source $\hat{X}_e$ for $e \in E$. Consider a source realization $(\hat{\mathbf{x}}_\mathbf{s}, \hat{\mathbf{x}}_\mathbf{e})$ in $\hat{\mathcal{I}}$ "corresponding" to $\mathbf{x} = \{x_s\}$: namely, for $s \in S$ set $\hat{\mathbf{x}}_\mathbf{s}(s) = x_s$ and set $\hat{\mathbf{x}}_\mathbf{e}$ to be any complementary source realization.

For source realization $\mathbf{x}$ of $\mathcal{I}$, let $\mathbf{x}_e$ be the realization of $X_e$ transmitted on edge $e$ in the execution of the network code $(\mathcal{F}, \mathcal{X})$. By our definitions, it holds that for any edge $e \in E$, $\bar{f}_e(\hat{\mathbf{x}}_\mathbf{s}) = \bar{f}_e(\mathbf{x}) = \mathbf{x}_e$. It follows that the realization of $\hat{X}_B(e) = \hat{X}_e + \bar{f}_e(\hat{X}_1, \ldots, \hat{X}_{|S|})$ is $\hat{\mathbf{x}}_\mathbf{e}(e) + \bar{f}_e(\hat{\mathbf{x}}_\mathbf{s}) = \hat{\mathbf{x}}_\mathbf{e}(e) + \mathbf{x}_e$. In addition, as we are assuming correct decoding on $\mathbf{x}$, for each terminal $t_i \in T$ of $\mathcal{I}$ it holds that $g_i(\mathbf{x}_{e'} : e' \in In(t_i)) = (x_s : b(s, t_i) = 1)$.

Consider a terminal $\hat{t}_e$ in $\hat{\mathcal{I}}$. The decoding procedure of $\hat{t}_e$ first computes for $e' \in In(e)$ the realization of $\hat{X}_B(e') + \hat{X}_{e'}$ which by the discussion above is exactly $\hat{\mathbf{x}}_\mathbf{e}(e') + \mathbf{x}_{e'} + \hat{\mathbf{x}}_\mathbf{e}(e') = \mathbf{x}_{e'}$. Then the decoder computes $f_e(\mathbf{x}_{e'} : e' \in In(e)) = \bar{f}_e(\mathbf{x}) = \bar{f}_e(\hat{\mathbf{x}}_\mathbf{s}) = \mathbf{x}_e$. Finally, the decoder computes the realization of $\hat{X}_B(e) + \bar{f}_e(\hat{X}_1, \ldots, \hat{X}_{|S|})$ which is $\hat{\mathbf{x}}_\mathbf{e}(e) + \mathbf{x}_e + \mathbf{x}_e = \hat{\mathbf{x}}_\mathbf{e}(e)$ which is exactly the information that the decoder needs.

Similarly, consider a terminal $\hat{t}_i$ in $\hat{\mathcal{I}}$ corresponding to a terminal $t_i \in T$ of $\mathcal{I}$. The decoding procedure of $\hat{t}_i$ first computes for $e' \in In(t_i)$ the realization of $\hat{X}_B(e') + \hat{X}_{e'}$ which by the discussion above is exactly $\mathbf{x}_{e'}$. Then the decoder computes $g_i(\mathbf{x}_{e'} : e' \in In(t_i)) = (x_s : b(s, t_i) = 1)$, which is exactly the information needed by $\hat{t}_i$.

Finally, consider the terminal $\hat{t}_{all}$. The decoding procedure of $\hat{t}_{all}$ computes for each $e \in E$ the realization of $\hat{X}_B(e) + \bar{f}_e(\hat{X}_1, \ldots, \hat{X}_{|S|})$ which is $\hat{\mathbf{x}}_\mathbf{e}(e) + \mathbf{x}_e + \mathbf{x}_e = \hat{\mathbf{x}}_\mathbf{e}(e)$ which again is exactly the information needed by $\hat{t}_{all}$.

All in all, we conclude that all terminals of $\hat{\mathcal{I}}$ decode correctly on source realization $(\hat{\mathbf{x}}_\mathbf{s}, \hat{\mathbf{x}}_\mathbf{e})$ corresponding to source realization $\mathbf{x}$ of $\mathcal{I}$ which allows correct decoding in $\mathcal{I}$. This implies that the instance $\hat{\mathcal{I}}$ is indeed $(\varepsilon, \hat{R}, \hat{c}_B, n)$ feasible.

$\hat{\mathcal{I}}$ **is** $(\varepsilon, \hat{R}, \hat{c}_B, n)$ **feasible implies that** $\mathcal{I}$ **is** $(\varepsilon, R, n)$ **feasible:**

Here, we assume that $\hat{\mathcal{I}}$ is $(\varepsilon, \hat{R}, \hat{c}_B, n)$ feasible with $\hat{c}_B$ as defined above. Thus, there exists an index code $(\hat{\mathcal{E}}_B, \hat{\mathcal{D}}) = (\hat{\mathcal{E}}_B, \{\hat{D}_{\hat{t}}\})$ for $\hat{\mathcal{I}}$ with block length $n$ and success probability at least $1 - \varepsilon$. In what follows we obtain a network code $(\mathcal{F}, \mathcal{X}) = \{(f_e, X_e)\} \cup \{g_t\}$ for $\mathcal{I}$. The key observation we use is that by our definition of $\hat{c}_B = \sum_{e \in E} \hat{R}_e$, the support $[2^{\hat{c}_B n}]$ of the encoding $\hat{\mathcal{E}}_B$ is exactly the size of the (product of) the supports of the source variables $\{\hat{X}_e\}$ in $\hat{\mathcal{I}}$. The implications of this observation are described below.

We start with some notation. For each realization $\hat{\mathbf{x}}_\mathbf{s} = \{\hat{x}_s\}_{s \in S}$ of source information $\{\hat{X}_s\}$ in $\hat{\mathcal{I}}$, let $A_{\hat{\mathbf{x}}_\mathbf{s}}$ be the realizations $\hat{\mathbf{x}}_\mathbf{e} = \{\hat{x}_e\}_{e \in E}$ of $\{\hat{X}_e\}_{e \in E}$ for which all terminals decode $(\hat{\mathbf{x}}_\mathbf{s}, \hat{\mathbf{x}}_\mathbf{e})$ correctly. That is, if we use the term "good" to refer to any source realization pair $(\hat{\mathbf{x}}_\mathbf{s}, \hat{\mathbf{x}}_\mathbf{e})$ for which all terminals decode correctly $(\hat{X}_s, \hat{X}_e) = (\hat{\mathbf{x}}_\mathbf{s}, \hat{\mathbf{x}}_\mathbf{e})$, then

$$A_{\hat{\mathbf{x}}_\mathbf{s}} = \{\hat{\mathbf{x}}_\mathbf{e} \mid \text{the pair } (\hat{\mathbf{x}}_\mathbf{s}, \hat{\mathbf{x}}_\mathbf{e}) \text{ is good}\}.$$

*Claim 1:* For any given $\sigma \in [2^{\hat{c}_B n}]$ and any $\hat{\mathbf{x}}_\mathbf{s}$, there is at most one $\hat{\mathbf{x}}_\mathbf{e} \in A_{\hat{\mathbf{x}}_\mathbf{s}}$ for which $\hat{\mathcal{E}}_B(\hat{\mathbf{x}}_\mathbf{s}, \hat{\mathbf{x}}_\mathbf{e}) = \sigma$.

*Proof:* Let $\hat{\mathbf{x}}_\mathbf{s} = \{\hat{x}_s\}_{s \in S}$ be a realization of the source information $\{\hat{X}_s\}_{s \in S}$. Treat the broadcasted value $\hat{\mathcal{E}}_B(\hat{\mathbf{x}}_\mathbf{s}, \hat{\mathbf{x}}_\mathbf{e})$ as a function of $\hat{\mathbf{x}}_\mathbf{e}$. Namely, set $\hat{\mathcal{E}}_B(\hat{\mathbf{x}}_\mathbf{s}, \hat{X}_e) = \hat{\mathcal{E}}_{\hat{\mathbf{x}}_\mathbf{s}}(\hat{X}_e)$.

Now, for any $\hat{\mathbf{x}}_\mathbf{s}$ and any $\hat{\mathbf{x}}_\mathbf{e} \in A_{\hat{\mathbf{x}}_\mathbf{s}}$, it holds that terminal $\hat{t}_{all}$ will decode correctly given the realization of the "has" set $\hat{H}_{\hat{t}_{all}} = \hat{\mathbf{x}}_\mathbf{s}$ and the broadcasted information $\hat{X}_B$ via $\hat{\mathcal{E}}_B$. Namely, $\hat{D}_{\hat{t}_{all}}(\hat{\mathcal{E}}_{\hat{\mathbf{x}}_\mathbf{s}}(\hat{\mathbf{x}}_\mathbf{e}), \hat{\mathbf{x}}_\mathbf{s}) = \hat{\mathbf{x}}_\mathbf{e}$. We now show (by means of contradiction) that the function $\hat{\mathcal{E}}_{\hat{\mathbf{x}}_\mathbf{s}}(\hat{\mathbf{x}}_\mathbf{e})$ obtains different values for different $\hat{\mathbf{x}}_\mathbf{e} \in A_{\hat{\mathbf{x}}_\mathbf{s}}$. This will suffice to prove our assertion.

Suppose that there are two values $\hat{\mathbf{x}}_\mathbf{e} \neq \hat{\mathbf{x}}'_\mathbf{e}$ in $A_{\hat{\mathbf{x}}_\mathbf{s}}$ such that $\hat{\mathcal{E}}_{\hat{\mathbf{x}}_\mathbf{s}}(\hat{\mathbf{x}}_\mathbf{e}) = \hat{\mathcal{E}}_{\hat{\mathbf{x}}_\mathbf{s}}(\hat{\mathbf{x}}'_\mathbf{e})$. This implies that $\hat{\mathbf{x}}_\mathbf{e} = \hat{D}_{\hat{t}_{all}}(\hat{\mathcal{E}}_{\hat{\mathbf{x}}_\mathbf{s}}(\hat{\mathbf{x}}_\mathbf{e}), \hat{\mathbf{x}}_\mathbf{s}) = \hat{D}_{\hat{t}_{all}}(\hat{\mathcal{E}}_{\hat{\mathbf{x}}_\mathbf{s}}(\hat{\mathbf{x}}'_\mathbf{e}), \hat{\mathbf{x}}_\mathbf{s}) = \hat{\mathbf{x}}'_\mathbf{e}$, which gives a contradiction. ∎

*Claim 2:* There exists a $\sigma \in [2^{\hat{c}_B n}]$ such that at least a $(1 - \varepsilon)$ fraction of source realizations $\hat{\mathbf{x}}_\mathbf{s}$ satisfy $\hat{\mathcal{E}}_B(\hat{\mathbf{x}}_\mathbf{s}, \hat{\mathbf{x}}_\mathbf{e}) = \sigma$ for some $\hat{\mathbf{x}}_\mathbf{e} \in A_{\hat{\mathbf{x}}_\mathbf{s}}$.

*Proof:* Consider a random value $\sigma$ chosen uniformly from $[2^{\hat{c}_B n}]$. For any partial source realization $\hat{\mathbf{x}}_\mathbf{s}$, the probability that there exists a realization $\hat{\mathbf{x}}_\mathbf{e} \in A_{\hat{\mathbf{x}}_\mathbf{s}}$ for which $\hat{\mathcal{E}}_B(\hat{\mathbf{x}}_\mathbf{s}, \hat{\mathbf{x}}_\mathbf{e}) = \sigma$ is at least $|A_{\hat{\mathbf{x}}_\mathbf{s}}|/2^{\hat{c}_B n}$. This follows by Claim 1, since for every $\hat{\mathbf{x}}_\mathbf{e} \in A_{\hat{\mathbf{x}}_\mathbf{s}}$ it holds that $\hat{\mathcal{E}}_B(\hat{\mathbf{x}}_\mathbf{s}, \hat{\mathbf{x}}_\mathbf{e})$ is distinct. Hence, the expected number of source realizations $\hat{\mathbf{x}}_\mathbf{s}$ for which there exists a realization $\hat{\mathbf{x}}_\mathbf{e} \in A_{\hat{\mathbf{x}}_\mathbf{s}}$ with $\hat{\mathcal{E}}_B(\hat{\mathbf{x}}_\mathbf{s}, \hat{\mathbf{x}}_\mathbf{e}) = \sigma$ is at least

$$\frac{\sum_{\hat{\mathbf{x}}_\mathbf{s}} |A_{\hat{\mathbf{x}}_\mathbf{s}}|}{2^{\hat{c}_B n}} \geq \frac{(1-\varepsilon)2^{n(\sum_{s \in S} \hat{R}_s + \sum_{e \in E} \hat{R}_e)}}{2^{\hat{c}_B n}}.$$

We use here the fact that the total number of source realizations $(\hat{\mathbf{x}}_\mathbf{s}, \hat{\mathbf{x}}_\mathbf{e})$ for which the index code $(\hat{\mathcal{E}}_B, \hat{\mathcal{D}})$ succeeds is exactly $\sum_{\hat{\mathbf{x}}_\mathbf{s}} |A_{\hat{\mathbf{x}}_\mathbf{s}}|$, which by the $\varepsilon$ error assumption is at least $(1 - \varepsilon)2^{n(\sum_{s \in S} \hat{R}_s + \sum_{e \in E} \hat{R}_e)}$.

Since $\hat{c}_B = \sum_{e \in E} \hat{R}_e$,

$$\frac{(1-\varepsilon)2^{n(\sum_{s \in S} \hat{R}_s + \sum_{e \in E} \hat{R}_e)}}{2^{\hat{c}_B n}} = (1-\varepsilon)2^{n(\sum_{s \in S} \hat{R}_s)},$$

which, in turn, is exactly the size of a $(1 - \varepsilon)$ fraction of all partial source realizations $\hat{\mathbf{x}}_\mathbf{s}$.

We conclude that there is a $\sigma \in [2^{\hat{c}_B n}]$ which "behaves" at least as well as expected, namely a value of $\sigma$ that satisfies the requirements in the assertion. ∎

We will now define the encoding functions of $(\mathcal{F}, \mathcal{X})$ for the network code instance $\mathcal{I}$. Specifically, we need to define the encoding functions $\{f_e\}$ and the decoding functions $\{g_t\}$ for the edges $e$ in $E$ and terminals $t$ in the terminal set $T$ of $\mathcal{I}$. We start by formally defining the functions. We then prove that they are an $(\varepsilon, R, n)$ feasible network code for $\mathcal{I}$.

Let $\sigma$ be the value specified in Claim 2, let $A_\sigma$ be the set of partial realizations $\hat{\mathbf{x}}_\mathbf{s}$ for which there exists a realization $\hat{\mathbf{x}}_\mathbf{e} \in A_{\hat{\mathbf{x}}_\mathbf{s}}$ with $\hat{\mathcal{E}}_B(\hat{\mathbf{x}}_\mathbf{s}, \hat{\mathbf{x}}_\mathbf{e}) = \sigma$. By Claim 2, the size of $A_\sigma$ is at least $(1 - \varepsilon)2^{n(\sum_{s \in S} \hat{R}_s)} = (1 - \varepsilon)2^{n(\sum_{s \in S} R_s)}$.

For $e \in E$ let

$$f_e : \left[2^{n \sum_{e' \in In(e)} c_{e'}}\right] \to [2^{nc_e}]$$

be the function that takes as input the random variables $(X_{e'} : e' \in In(e))$ and outputs $X_e = \hat{D}_{\hat{t}_e}(\sigma, (X_{e'} : e' \in In(e)))$. Here, we consider $X_{e'}$ for $e' \in E$ to be a random variable of support $[2^{c_{e'} n}]$.

For terminals $t_i \in T$ in $\mathcal{I}$ let

$$g_{t_i} : \left[2^{n \sum_{e' \in In(t_i)} c_{e'}}\right] \to \left[2^{n \sum_{s \in S : b(s, t_i) = 1} R_s}\right]$$

be the function that takes as input the random variables $(X_{e'} : e' \in In(t_i))$ and outputs $\hat{D}_{\hat{t}_i}(\sigma, (X_{e'} : e' \in In(t_i)))$.

We will now show that the network code defined above decodes correctly with probability $1 - \varepsilon$. Consider any rate $R = (R_1, \ldots, R_{|S|})$ realization of the source information in $\mathcal{I}$: $\mathbf{x} = \{x_s\}$. Consider the source information $\hat{\mathbf{x}}_\mathbf{s}$ of $\hat{\mathcal{I}}$ corresponding to $\mathbf{x}$, namely let $\hat{\mathbf{x}}_\mathbf{s} = \mathbf{x}$. Assume that $\hat{\mathbf{x}}_\mathbf{s} \in A_\sigma$. Using Claim 2, let $\hat{\mathbf{x}}_\mathbf{e}$ be the realization of source information $\{\hat{X}_e\}$ in $\hat{\mathcal{I}}$ for which $\hat{\mathcal{E}}_B(\hat{\mathbf{x}}_\mathbf{s}, \hat{\mathbf{x}}_\mathbf{e}) = \sigma$. Recall that, by our definitions, all terminals of $\hat{\mathcal{I}}$ will decode correctly given source realization $(\hat{\mathbf{x}}_\mathbf{s}, \hat{\mathbf{x}}_\mathbf{e})$. For $s \in S$, let $\hat{\mathbf{x}}_\mathbf{s}(s) = x_s$ be the entry in $\hat{\mathbf{x}}_\mathbf{s}$ that corresponds to $\hat{X}_s$. For $e \in E$, let $\hat{\mathbf{x}}_\mathbf{e}(e)$ be the entry in $\hat{\mathbf{x}}_\mathbf{e}$ that corresponds to $\hat{X}_e$.

We show by induction on the topological order of $G$ that for source information $\mathbf{x}$ the evaluation of $f_e$ in the network code above results in the value $x_e$ which is equal to $\hat{\mathbf{x}}_\mathbf{e}(e)$. For the base case, consider an edge $e = (u, v)$ in which $u$ is a source with no incoming edges. In that case, by our definitions, the information $x_e$ on edge $e$ equals $f_e(x_s) = \hat{D}_{\hat{t}_e}(\sigma, x_s) = \hat{D}_{\hat{t}_e}(\hat{\mathcal{E}}_B(\hat{\mathbf{x}}_\mathbf{s}, \hat{\mathbf{x}}_\mathbf{e}), \hat{\mathbf{x}}_\mathbf{s}(s)) = \hat{\mathbf{x}}_\mathbf{e}(e)$. Here, the last equality follows from the fact that the index code $(\hat{\mathcal{E}}_B, \hat{\mathcal{D}})$ succeeds on source realization $(\hat{\mathbf{x}}_\mathbf{s}, \hat{\mathbf{x}}_\mathbf{e})$, and thus all terminals (and, in particular, terminal $\hat{t}_e$) decode correctly.

In general, consider an edge $e = (u, v)$ with incoming edges $e' \in In(e)$. In that case, by our definitions, the information $x_e$ on edge $e$ equals $f_e(x_{e'} : e' \in In(e))$. However, by induction, each $x_{e'}$ for which $e' \in In(e)$ satisfies $x_{e'} = \hat{\mathbf{x}}_\mathbf{e}(e')$. Thus $x_e = \hat{D}_{\hat{t}_e}(\sigma, (x_{e'} : e' \in In(e))) = \hat{D}_{\hat{t}_e}(\hat{\mathcal{E}}_B(\hat{\mathbf{x}}_\mathbf{s}, \hat{\mathbf{x}}_\mathbf{e}), (\hat{\mathbf{x}}_\mathbf{e}(e') : e' \in In(e))) = \hat{\mathbf{x}}_\mathbf{e}(e)$. As before, the last equality follows from the fact that the index code $(\hat{\mathcal{E}}_B, \hat{\mathcal{D}})$ succeeds on source realization $(\hat{\mathbf{x}}_\mathbf{s}, \hat{\mathbf{x}}_\mathbf{e})$.

Finally, we address the value of the decoding functions $g_t$. By definition, the outcome of $g_t$ is exactly $\hat{D}_{\hat{t}_i}(\sigma, (x_{e'} : e' \in In(t_i))) = \hat{D}_{\hat{t}_i}(\hat{\mathcal{E}}_B(\hat{\mathbf{x}}_\mathbf{s}, \hat{\mathbf{x}}_\mathbf{e}), (\hat{\mathbf{x}}_\mathbf{e}(e') : e' \in In(t_i))) = (\hat{\mathbf{x}}_\mathbf{s}(s) : b(s, t_i) = 1) = (x_s : b(s, t_i) = 1)$. As before, we use the inductive argument stating that $x_{e'} = \hat{\mathbf{x}}_\mathbf{e}(e')$, and the fact that the index code $(\hat{\mathcal{E}}_B, \hat{\mathcal{D}})$ succeeds on source realization $(\hat{\mathbf{x}}_\mathbf{s}, \hat{\mathbf{x}}_\mathbf{e})$, and thus all terminals (and, in particular, terminal $\hat{t}_i$) decode correctly. This suffices to show that the proposed network code $(\mathcal{F}, \mathcal{X})$ succeeds with probability $1-\varepsilon$ on source input of rate tuple $R$. We have presented correct decoding for $\mathcal{I}$ when $\mathbf{x} = \hat{\mathbf{x}}_\mathbf{s} \in A_\sigma$, and shown that $|A_\sigma| \geq (1-\varepsilon) 2^{n(\sum_{s \in S} R_s)}$. Therefore, we have shown correct decoding for $\mathcal{I}$ with probability at least $(1-\varepsilon)$. ∎

## V. CAPACITY REGIONS

In certain cases, our connection between network and index coding presented in Theorem 1 implies a tool for determining the network coding capacity via the capacity of index coding instances. Below, we present such a connection in the case of *collocated* sources (i.e., for network coding instances in which all the sources are collocated at a single node in the network). Similar results can be obtained for "super source" networks (studied in, e.g., [19], [20]). We discuss general network coding instances in Section VI.

*Corollary 1:* For any instance to the network coding problem $\mathcal{I}$ where all sources are collocated, one can efficiently construct an instance to the index coding problem $\hat{\mathcal{I}}$ and an integer $\hat{c}_B$ such that for any rate tuple $R$: $R$ is in the capacity region of $\mathcal{I}$ iff $(\hat{R}, \hat{c}_B)$ is in the capacity region of $\hat{\mathcal{I}}$. Here, the rate vector $\hat{R}$ for $\hat{\mathcal{I}}$ can be efficiently constructed from $R$.

*Proof:* Let $\mathcal{I}$ be an instance to the network coding problem and let $R$ be any rate tuple. The instance $\hat{\mathcal{I}}$, the rate tuple $\hat{R}$ and the integer $\hat{c}_B$ are obtained exactly as presented in Theorem 1. We now show that any $R$ is in the capacity region of $\mathcal{I}$ iff $(\hat{R}, \hat{c}_B)$ is in the capacity region of $\hat{\mathcal{I}}$.

**From network coding to index coding:** Suppose that $R$ is in the capacity region of the network coding instance $\mathcal{I}$. Namely, for any $\varepsilon > 0$, any $\delta > 0$, and source rates $R(1-\delta) = (R_1(1-\delta), \ldots, R_{|S|}(1-\delta))$, there exists a network code with a certain block length $n$ that satisfies $\mathcal{I}$ with error probability $\varepsilon$. As shown in the proof of the first direction of Theorem 1, this network code can be efficiently mapped to an index code for $\hat{\mathcal{I}}$ of block length $n$, broadcast rate equal to $\hat{c}_B n$, error probability $\varepsilon$ and source rates $\hat{R}_\delta = (\{\hat{R}_s(1-\delta)\}_{s \in S}, \{\hat{R}_e\}_{e \in E})$. Therefore, for any $\varepsilon > 0$ and any $\delta > 0$, there exists a block length $n$ for which $\hat{\mathcal{I}}$ is $(\varepsilon, \hat{R}_\delta, \hat{c}_B, n)$ feasible, and thus $(\hat{R}, \hat{c}_B)$ is in the capacity region of $\hat{\mathcal{I}}$.

**From index coding to network coding:** Suppose that $(\hat{R}, \hat{c}_B)$ is in the capacity region of $\hat{\mathcal{I}}$. Recall that $\hat{R} = (\{\hat{R}_s\}_{s \in S}, \{\hat{R}_e\}_{e \in E})$ and $\hat{c}_B = \sum_{e \in E} \hat{R}_e$. Therefore, for any $\varepsilon > 0$ and any $\delta \geq 0$ there exists an index code with a certain block length $n$ and error probability $\varepsilon$ such that $\hat{\mathcal{I}}$ is $(\varepsilon, \hat{R}(1-\delta), \hat{c}_B, n)$-feasible. Note that we cannot readily use the proof of the second direction of Theorem 1 to map this index code into an network code for $\mathcal{I}$. That is because this map requires that $\hat{c}_B$ be equal to the sum of rates of random variables in $\hat{\mathcal{I}}$ that correspond to edges in $E$, namely that $\hat{c}_B = (1-\delta)\sum_{e \in E} \hat{R}_e$. However, in our setting we have $\hat{c}_B = \sum_{e \in E} \hat{R}_e$. This (small) slackness will not allow the proof of Theorem 1 to go through. Instead, we proceed by first stating a claim similar to Claim 2, which will allow us to prove our results in the setting in which all sources in $\mathcal{I}$ are collocated. The proof of Claim 3 appears at the end of this section. Throughout, we use the notation set in the proof of Theorem 1.

*Claim 3:* There exists a set $\Sigma \subset [2^{\hat{c}_B n}]$ of cardinality
$$|\Sigma| = n \log(4/3)(1-\delta)(\sum_{s \in S} \hat{R}_s) 2^{n\delta \sum_{e \in E} \hat{R}_e}$$
such that least a $(1-2\varepsilon)$ fraction of source realizations $\hat{\mathbf{x}}_\mathbf{s}$ satisfy $\hat{\mathcal{E}}_B(\hat{\mathbf{x}}_\mathbf{s}, \hat{\mathbf{x}}_\mathbf{e}) = \sigma$ for some $\hat{\mathbf{x}}_\mathbf{e} \in A_{\hat{\mathbf{x}}_\mathbf{s}}$ and some $\sigma \in \Sigma$.

Assuming Claim 3 we will now define the encoding and decoding functions for the network coding instance $\mathcal{I}$. Suppose that all the sources $s \in S$ are collocated at a single node that we call the source node. For each source realization $\mathbf{x}_\mathbf{s}$, the source node checks whether there exists $\hat{\mathbf{x}}_\mathbf{e} \in A_{\hat{\mathbf{x}}_\mathbf{s}}$ and some $\sigma_{\mathbf{x}_\mathbf{s}} \in \Sigma$ such that $\hat{\mathcal{E}}_B(\hat{\mathbf{x}}_\mathbf{s}, \hat{\mathbf{x}}_\mathbf{e}) = \sigma_{\mathbf{x}_\mathbf{s}}$ (Case A) or not (Case B).

In Case A, the network code operates in two phases. During the first phase, the source node sends an overhead message to all the nodes in the network[2] revealing the value of $\sigma_{\mathbf{x}_\mathbf{s}}$ using at most $\log |\Sigma|$ bits (by Claim 3). The rate needed to convey the overhead message vanishes for arbitrarily small values of $\delta$. In the second phase, we implement the network code described in the second direction of the proof of Theorem 1 (with $\sigma = \sigma_{\mathbf{x}_\mathbf{s}}$). The source $\mathbf{x}_\mathbf{s}$ is transmitted through the network by sending on edge $e$ the message $X_e = \hat{D}_{\hat{t}_e}(\sigma_{\mathbf{x}_\mathbf{s}}, (X_{e'} : e' \in In(e)))$ and each terminal $t_i$ implementing the decoding function $\hat{D}_{\hat{t}_i}(\sigma_{\mathbf{x}_\mathbf{s}}, (X_{e'} : e' \in In(t_i)))$. The total block length used is $n(1+\delta')$ for $\delta' = \log |\Sigma|/n$ that tends to zero as $\delta$ tends to zero. It is not hard to verify (based on the proof of Theorem 1) that such encoding/decoding functions for $\mathcal{I}$ will allow successful decoding when the source realization for $\mathcal{I}$ is $\mathbf{x}_\mathbf{s}$.

In Case B, we allow the network to operate arbitrarily, and consider this case as an error. Claim 3 implies that Case B will happen with probability at most $2\epsilon$. Therefore, for $R = (\hat{R}_{s_1}, \ldots, \hat{R}_{s_{|S|}})$ the network coding instance $\mathcal{I}$ is $(2\varepsilon, \frac{R(1-\delta)}{1+\delta'}, n(1+\delta'))$ feasible for $\delta' = \log |\Sigma|/n$. As $2\varepsilon$ tends to zero with $\varepsilon$ tending to zero, and similarly $\delta'$ tends to zero as $\delta$ tends to zero, we conclude that $\mathcal{I}$ is $R$-feasible. ∎

We now present the proof of Claim 3.

*Proof:* (Claim 3) Consider the elements $\hat{\mathbf{x}}_\mathbf{s}$ for which $|A_{\hat{\mathbf{x}}_\mathbf{s}}|$ is at least of size $2^{n(1-\delta)\hat{c}_B - 1}$. Recall that $\hat{c}_B =$

---
[2] Any node that cannot be reached by a directed path from the source node can be set to remain inactive (not transmit any message) without altering the capacity region of the network.

$\sum_{e \in E} \hat{R}_e$. Denote these elements $\hat{\mathbf{x}}_\mathbf{s}$ by the set $A$. Notice that
$$|A| \geq (1 - 2\varepsilon) \cdot 2^{n(1-\delta)\sum_{s \in S} \hat{R}_s}.$$

Otherwise the total error in the index code we are considering is greater than $\varepsilon$, which is a contradiction to our assumption.

Let $\Sigma'$ be a subset of $[2^{n\hat{c}_B}]$ of cardinality $|\Sigma'| = 2^{\delta n \hat{c}_B}$ chosen uniformly at random (i.e., each element of $\Sigma'$ is i.i.d. uniform from $[2^{n\hat{c}_B}]$). For $\hat{\mathbf{x}}_\mathbf{s} \in [2^{n(1-\delta)\sum_{s \in S} \hat{R}_s}]$ define the binary random variable $Z_{\hat{\mathbf{x}}_\mathbf{s}}$, such that $Z_{\hat{\mathbf{x}}_\mathbf{s}} = 1$ whenever there exist $\hat{\mathbf{x}}_\mathbf{e} \in A_{\hat{\mathbf{x}}_\mathbf{s}}$ and $\sigma \in \Sigma'$ such that $\hat{\mathcal{E}}_B(\hat{\mathbf{x}}_\mathbf{s}, \hat{\mathbf{x}}_\mathbf{e}) = \sigma$, and $Z_{\hat{\mathbf{x}}_\mathbf{s}} = 0$ otherwise.

Using Claim 1, we have for any $\hat{\mathbf{x}}_\mathbf{s} \in A$ that
$$\Pr(Z_{\hat{\mathbf{x}}_\mathbf{s}} = 1) = 1 - \left(1 - \frac{|A_{\hat{\mathbf{x}}_\mathbf{s}}|}{2^{n\hat{c}_B}}\right)^{|\Sigma'|}$$
$$\geq 1 - \left(1 - \frac{2^{n(1-\delta)\hat{c}_B - 1}}{2^{n\hat{c}_B}}\right)^{|\Sigma'|}$$
$$= 1 - \left(1 - \frac{1}{2|\Sigma'|}\right)^{|\Sigma'|} > \frac{1}{4}.$$

We say that the $\hat{\mathbf{x}}_\mathbf{s} \in A$ is *covered* by $\Sigma'$ if $Z_{\hat{\mathbf{x}}_\mathbf{s}} = 1$. It suffices to cover all $\hat{\mathbf{x}}_\mathbf{s} \in A$ in order to satisfy our assertion.

In expectation, $\Sigma'$ covers at least $\frac{1}{4}$ of the elements in $A$. Using the same averaging argument as in Claim 2, it follows that there exists a choice for the set $\Sigma'$ that covers $\frac{1}{4}$ of the elements in $A$. By removing these covered values of $\hat{\mathbf{x}}_\mathbf{s}$ and repeating on the remaining elements in $A$ in a similar manner iteratively, we can cover all the elements of $A$. Specifically, iterating $\log |A| / \log(4/3)$ times (each time with a new $\Sigma'$) it is not hard to verify that all elements of $A$ will eventually be covered. Taking $\Sigma$ to be the union of all $\Sigma'_i$ obtained in iteration $i$, we conclude our assertion. ∎

## VI. CONCLUSIONS

In this work, we addressed the equivalence between the network and index coding paradigms. Following the line of proof presented in [17] for a restricted equivalence in the case of linear encoding, we present an equivalence for general (not necessarily linear) encoding functions. Our results show that the study and understanding of the index coding paradigm implies a corresponding understanding of the network coding paradigm.

Although our connection between network and index coding is very general it does not directly imply a tool for determining the network coding capacity region as defined in Section II for general network coding instances. Indeed, as mentioned in the proof of Corollary 1 for collocated sources, a naive attempt to reduce the problem of determining whether a certain rate vector $R$ is in the capacity region of a network coding instance $\mathcal{I}$ to the problem of determining whether a corresponding rate vector $\hat{R}$ is in the capacity region of an index coding instance $\hat{\mathcal{I}}$, shows that a stronger, more robust connection between index and network coding is needed. A connection which allows some flexibility in the value of the broadcast rate $\hat{c}_B$. Such a connection is subject to future studies.

Recently, it has been shown [20], [21] that certain intriguing open questions in the context of network coding are well understood in the context of index coding (or the so-called "super-source" setting of network coding). These include the question of "zero-vs-$\varepsilon$ error: "What is the maximum loss in rate when insisting on zero error communication as opposed to vanishing decoding error?" [19], [20]; the "edge removal" question: "What is the maximum loss in communication rate experienced from removing an edge of capacity $\delta > 0$ from a given network?" [22], [23]; and the following question regarding the independence of source information: "What is the maximum loss in rate when comparing the communication of source information that is "almost" independent to that of independent source information?" [21].

At first, it may seem that the equivalence presented in this work implies a full understanding of the open questions above in the context of network coding. Although this may be the case, a naive attempt to use our results with those presented in [20], [21] again shows the need of a stronger connection between index and network coding that (as above) allows some flexibility in the value of $\hat{c}_B$.

## VII. ACKNOWLEDGEMENTS

S. El Rouayheb would like to thank Prof. H. Vincent Poor for his valuable feedback and continuous support, and Curt Schieler for interesting discussions on the index coding problem.